\documentstyle[11pt,epsf]{article}
\input psfig
\def\beq{\begin{eqnarray}}
\def\eeq{\end{eqnarray}}
\def\bea{\begin{eqnarray*}}
\def\eea{\end{eqnarray*}}

\def\centeron#1#2{{\setbox0=\hbox{#1}\setbox1=\hbox{#2}\ifdim
\wd1>\wd0\kern.5\wd1\kern-.5\wd0\fi
\copy0\kern-.5\wd0\kern-.5\wd1\copy1\ifdim\wd0>\wd1
\kern.5\wd0\kern-.5\wd1\fi}}
\def\ltap{\;\centeron{\raise.35ex\hbox{$<$}}{\lower.65ex\hbox{$\sim$}}\;}
\def\gtap{\;\centeron{\raise.35ex\hbox{$>$}}{\lower.65ex\hbox{$\sim$}}\;}

\def\singleandthirdspaced{\baselineskip=\normalbaselineskip\multiply
    \baselineskip by 130\divide\baselineskip by 100}

\newcommand{\newc}{\newcommand}
\newc{\qbar}{{\overline q}}
\newc{\Kahler}{K\"ahler }
\newc{\deltaGS}{\delta_{\rm GS}}
\begin{document}
\begin{titlepage}
\begin{flushright}
{\large hep-th/0501214 \\ SCIPP-2004/21\\
}
\end{flushright}

\vskip 1.2cm

\begin{center}

{\LARGE\bf Branches of the Landscape}

\vskip 1.4cm

{\large  M. Dine, D. O'Neil and Z. Sun}
\\
\vskip 0.4cm
{\it Santa Cruz Institute for Particle Physics,
     Santa Cruz CA 95064  } \\

\vskip 4pt

\vskip 1.5cm

\begin{abstract}
With respect to the question
of supersymmetry breaking, there are three branches of
the flux landscape.  On one of these, if one requires small
cosmological constant, supersymmetry
breaking is predominantly at the fundamental scale; on another,
the distribution is roughly flat on a logarithmic scale; on the third,
the preponderance of vacua are at very low scale.  A priori, as we
will explain, one can say little about the first branch.
The vast majority of these states are not accessible even to crude, approximate
analysis.  On the other two branches one can hope to do better.
But as a result of the lack of access to branch one, and our poor
understanding of cosmology, we can at best conjecture
about whether string theory predicts low energy supersymmetry or not.  If we
hypothesize that are on branch two or three, distinctive
predictions may be possible.  We comment of the status of naturalness
within the landscape, deriving, for example, the statistics of the first
branch from simple effective field theory reasoning.
\end{abstract}

\end{center}

\vskip 1.0 cm

\end{titlepage}
\setcounter{footnote}{0} \setcounter{page}{2}
\setcounter{section}{0} \setcounter{subsection}{0}
\setcounter{subsubsection}{0}

\singleandthirdspaced

\section{Introduction:  Landscape Phenomenology}

The cosmological constant problem is one of the greatest puzzles confronting
particle theory and theories of gravity.    Until very recently, string theory
seemed to offer no solution to this problem.  Now, however, there is evidence
that the vast array of possible fluxes in string theory leads to a discretuum
of metastable
states\cite{Bousso:2000xa,Feng:2000if,Becker:1996gj,deBoer:2001px,
Giddings:2001yu,Acharya:2002kv,Kachru:2003aw,Susskind:2003kw}.
The number may well be large enough that a non-trivial
subset have cosmological constant of the order observed.

This strongly suggests that string theory can accommodate the observed cosmological
constant.  While one might hope that some cosmological
effect would pick out the observed vacuum uniquely,
it seems more likely that explaining the cosmological constant in such a framework
will require invoking
anthropic considerations. Absent any real understanding of string cosmology,
adopting the hypothesis
that somehow the universe, in its history,
explores a large fraction of these states, in a more or less democratic
fashion, predicts a cosmological constant in roughly the right
range\cite{Weinberg:1988cp,Garriga:1999bf,Weinberg:2000yb}.

The question is:  having adopted the hypotheses that the landscape
exists and that the universe samples all of these states, can we explain or predict
anything further?  Doing so requires an understanding of the statistics
of string vacuum states\cite{Douglas:2003um,Ashok:2003gk,
Douglas:2004kp,Denef:2004ze,
Denef:2004cf}.  There has already been significant progress in this
direction, and we can begin to address certain questions.  Perhaps the most
obvious is the origin of the hierarchy between the weak scale and the Planck (or
other large) scales.  The familiar proposals:  low energy supersymmetry, with
some sort of dynamical supersymmetry breaking, and technicolor have realizations
in the landscape which are at least partially understood.  So it is natural to
ask whether the landscape might predict low energy supersymmetry, perhaps
with some specific breaking scale and pattern of soft masses, or something
like technicolor (or, probably equivalently, warped geometry), or whether
something totally different might arise.  Perhaps the most troubling alternative
suggested by the landscape is that the solution is similar to that of the cosmological
constant -- there is simply a huge number of states, and anthropic considerations
pick out the observed hierarchy, with no other low energy consequences.

In \cite{Banks:2003es}, a critique of this program was offered, and the
very existence of the landscape was questioned (a more extensive discussion
appears in \cite{Banks:2004xh}, and some possible counterarguments
in \cite{Freivogel:2004rd}), but it was also suggested
that if the landscape does exist, low energy supersymmetry would be a plausible
outcome.  This statement was met by some
skepticism\cite{Douglas:2004qg,Susskind:2004uv}; it was argued that
supersymmetry would necessarily be broken at the Planck
scale. These arguments, while also plausible,
relied on assumptions whose validity has not been established.  A more
refined description of how low energy
supersymmetry might emerge
was presented in \cite{Dine:2004is}.  These
authors distinguished three branches of the landscape.
In the case of Type IIB theories compactified on orientifolds of Calabi-Yau spaces
(or F theory compactifications on Calabi-Yau four-folds), these correspond to:
\begin{enumerate}
\item  States with broken supersymmetry at tree level
\item  States with unbroken supersymmetry and $W \ne 0$ at tree level
\item  States with unbroken supersymmetry and $W =0$ at tree level.
\end{enumerate}
An extensive statistical analysis of
branch 2 was performed in \cite{Denef:2004ze}.  The statistics
of some simple models on branch three has been studied recently in \cite{DeWolfe:2004ns}.
More recently, in \cite{Denef:2004cf} the statistics of states on branch one with small scale of supersymmetry
breaking has recently been analyzed.

In this paper, we will attempt to understand the statistics and the physics of these
three branches further.  We begin by noting that, non-perturbatively,
there is no sharp distinction between the states on the different branches: superpotentials
are likely generated and supersymmetry broken on both branches two
and three. The real distinction between them lies in their
statistics.  If one considers the distribution of low energy lagrangians,
on branch one the measure is smooth in the regions where
$F$, the supersymmetry breaking scale and $\Lambda$ are small.  On branch
two, it is mildly singular in these regions; on branch three it is very singular.

We will discuss a number of features of
branch one.  We review why, in the classical approximation,
the number of stationary points of the supergravity potential is infinite,
and discuss the possible physics of the cutoff.
We give a simple, low energy effective action argument
which yields the statistics of \cite{Denef:2004cf}.  We explain why, on this branch, the vast
majority of the states are not accessible to analysis, so that it is not possible
to determine their statistics.    This means that, at present, one cannot decide, by a priori
theoretical arguments, whether this branch dominates, and, if this branch dominates,
one cannot make predictions.  We will give some reasons to conjecture that if this
branch dominates, it leads to phenomena similar either to technicolor theories or
to Randall Sundrum models.  We argue that there may already be phenomenological considerations
which rule out this possibility.

We then turn to branches two and three.  We want to ask:  if one adopts the hypothesis
that the universe is not on branch one, can one make predictions for quantities such
as the scale of supersymmetry breaking, and the pattern of soft breakings.
We will review some of what is known about branch two, and discuss the question
of computability.  Ref. \cite{DeWolfe:2004ns} considered models on branch three with
small numbers of moduli.  Another perspective on 
R symmetries was provided in \cite{Dienes:2004pi}.  We discuss the general problem of finding models with
vanishing $\langle W \rangle$, arguing that this is most likely connected with
discrete $R$ symmetries.  We explain that with discrete $R$ symmetries
statistical considerations will generally
lead to supersymmetric ground states with vanishing $W$.  We discuss in a well-known
model the statistical cost of such discrete symmetries and find that it is substantial.
But we note that a number of selection effects may still favor such vacua; moreover,
as suggested in\cite{DeWolfe:2004ns},
the statistical cost may not always be so high.

We conclude with our views of the prospects for prediction within the landscape.
The question of whether the theory predicts low energy supersymmetry, technicolor-like
theories, or nothing beyond the standard model Higgs seems to be an experimental
one, though there is some reason to think that the usual naturalness arguments -- even
in the landscape and even allowing possible
anthropic considerations -- will disfavor the last possibility.
On the supersymmetric branches, there is good reason to think that low energy
supersymmetry is favored.  Determining whether there are more
detailed predictions -- gauge mediation, anomaly mediation, or something completely
different, requires studies of the statistics of gauge groups beyond those which
have been done to date, but which may well be feasible.

\section{The Statistics of Branch One}

In string theory, we have become accustomed to the idea that the
more supersymmetry, the more control one has over the analysis of a problem.
As we will see, many features of branch three are particularly easy to
understand.  Ref. \cite{Denef:2004ze} studies the statistics
of branch two and some aspects of those of branch one.
Like KKLT, these authors study principally the Type IIB
theory on an orientifold of a Calabi-Yau space.  In this case, there is a superpotential
for the complex structure moduli in the supergravity approximation, but not for
the Kahler moduli.  Noting KKLT's observation
that the Kahler moduli are likely to be fixed
by non-perturbative effects,
they make the further simplification of ignoring the Kahler
moduli.  The most primitive aspect of
statistics is to simply count the number of stationary points
of the potential.  Here already there is a striking difference
between branches one and two; on the supersymmetric branch, the number
of stationary points is finite, while on the non-supersymmetric branch,
the number is divergent and it is necessary to introduce a cutoff.

In the limit in which the flux numbers are large, DD
convert sums
over fluxes to integrals over the moduli space. The
result has the form, in the supersymmetric case (branch two):
\beq
{\cal N}_{susy} = {L^{K/2} \over (K/2)!}{\cal N}_o.
\label{nsusy}
\eeq
${\cal N}_o$ is given by an expression of the form:
\beq
{\cal N}_o = \int d^{2m}z \int d^K N e^{-\vert X \vert^2 -\vert Z \vert^2
+ \vert Y_i \vert^2} \delta^{2m}(DW) \vert D^2 W \vert.
\label{nnaught}
\eeq
where
\beq
X= W  ~~~~ Y_i = D_{z_i} W  = F_{z_i}
\eeq
In the non-supersymmetric case, the term $\delta^{2m}(DW)$ is replaced
by $\delta^{2m}(V^\prime)$, and an appropriate determinant
We will not need the definition of $Z$; the main point is that it gives
a negative contribution to the exponential.
For supersymmetric vacua, Y vanishes, and the integral is finite.
This is not the case for non-supersymmetric vacua, and
Douglas and Denef argue that the integral is
divergent, and impose
a cutoff on the supersymmetry breaking scale ($F=Y=D_zW$), and argue
that it should not be too large.

While we believe that there may well be a physical cutoff, 
we do not have a convincing argument as to whether a physical cutoff exists.
Certainly one cannot calculate, even crudely, at large $F$;
as we will explain, the $\alpha^\prime$ expansion breaks down
badly once $F$ is comparable to the string scale.
We will also see that, for $F$ comparable to the string
scale, most of the would-be metastable states are highly unstable.
But while these arguments suggest that many would-be vacua are not
meaningful states of the theory, some number of states -- possibly infinite -- 
may still exist.
If there is a cutoff, there is no parameter which would make it small.  So, even
if we demand that states have small cosmological constant and are highly metastable,
most such states will be at the string scale.  The original
assumption that IIB orientifolds (or F theory on CY fourfolds) are representative
is thus not correct (or at least
not meaningful) on this
branch.  The typical states do not look like anything we know how to describe.
This means that we do not know how to do even the crudest counting on this branch.

\subsection{An Infinite Number of Non-Supersymmetric States}

We can demonstrate that the number of
non-supersymmetric stationary points of
the supergravity potential is infinite in a particular
class of flux models:  the IIB theory on $T_6/Z_2$\cite{Kachru:2002he}
The tadpole cancellation
condition reads:
\beq
-a^o d_o - a^{ij} d_{ij} + b_{ij}c^{ij} + b_o c^o = \ell^*
\eeq
for a fixed, positive integer $\ell^*$.  We can satisfy this condition by choosing:
\beq
b_o = -3N; c^o=N; c^{ij} =N \delta^{ij} = b_{ij}
\eeq
and $a^{ij}$, $a_o,d_o, d_{ij}$ numbers of order one which satisfy
the constraint.  Now $N$ can be taken arbitrarily large, so it would
appear that we have an infinite number of possible states.  We might
expect, however, that many of these are equivalent, related by various
modular transformations.  To establish that this is {\it not} the case,
we can do the following exercise.  In the large $N$ limit, the superpotential
takes the form:
\beq
W= N( -\phi \tau^3 +3 \phi \tau^2-3\tau + 3)
\eeq
We can then look for supersymmetric and non-supersymmetric solutions.
We have checked that there are no supersymmetric solutions, but there are
non-supersymmetric
ones:
\beq
\phi = {1 \over 3}(1 + i \sqrt{2}) ~~~~\tau = 1 + i \sqrt{2}.
\eeq
The energy of these solutions is:
\beq
V_o = 1.5 N^2
\eeq

Because the energy at the minimum of the potential is gauge invariant, and because
$N$ can be arbitrarily large, this result indicates that there are indeed an
infinite number of gauge-inequivalent vacuum states in without supersymmetry
in the supergravity approximation.

Note that the energy scales like $N^2$.  This is general for states with
large flux.  Correspondingly, the $F$ components of fields scale like $N$.
This will be important shortly when we discuss the $\alpha^\prime$ expansion.

These stationary points of the effective
action are unlikely to have any physical significance.  As we will explain
in the next section, the $\alpha^\prime$ expansion is invalid for them, and
the vast majority are very short lived.  

\subsection{The $\alpha^\prime$ Expansion and the Cutoff on the Non-Supersymmetric States}

It is easy to see that for
non-supersymmetric states, the $\alpha^\prime$ expansion already breaks down
for $F$-terms of order one.
In the $\alpha^\prime$ expansion, we know that there are
terms with additional derivatives of the complex structure ($z$) fields (derivatives
in the ``non-compact" directions.  In the effective
lagrangian, these terms arise from operators in superspace
with extra covariant derivatives, such as:
\beq
\int d^4 \theta D_\alpha z D^\alpha z \bar D_{\dot \alpha} z^* \bar D^{\dot \alpha} z^*.
\eeq
This includes $\vert F_z \vert^4$.

So $F_z$ is a measure of the reliability of the $\alpha^\prime$ expansion.
If all other quantities are roughly order one, it is necessary the $F_z$
be small in order to have any semblance of a controlled analysis.

\subsection{The Number of States at Small $F$}

At small $F$, one can hope to perform a self-consistent analysis.
Douglas and Denef have studied the number of states at small $F$.  Essentially
this involves evaluating the integral of eqns. \ref{nsusy},\ref{nnaught}, but with the delta function
for unbroken supersymmetry replaced by delta functions which enforces the minimum
conditions.  They find that the number of
non-supersymmetric states, with cosmological constant less than $\Lambda_o$ and
$\vert F \vert < F^*$ scales as:
\beq
{\cal N}(\Lambda<\Lambda_o,F<F^*) \approx \Lambda_o F^{*6}.
\eeq

\subsubsection{A Low Energy Understanding of the Douglas-Denef Results}

At first sight, the powers of $F^*$ found by Douglas and Denef may appear
surprising.  But a simple effective lagrangian analysis, coupled with a plausible
assumption about the distribution of couplings, yields the same results.
In the case that $\Lambda_o \ll F^{*2} \ll 1$,
the low energy effective theory is approximately a flat space, supersymmetric
one.
In this case, small supersymmetry breaking requires a light fermion
to provide the longitudinal mode of the gravitino.  In theories with chiral fields
only, this means that there must be at least one chiral field at scales well
below the fundamental scale.  It is certainly more probable to have one than several,
so the low energy theory consists of one chiral field (from the
complex structure moduli).  Call this field $z$, and define it so that the
minimum of its potential occurs at $z=0$ (to avoid any loss
of generality, we will, below, consider a very general class
of Kahler potentials).  Since, by assumption, $F$ is small,
the effective lagrangian is supersymmetric, so it is described by a superpotential
and a Kahler potential.

We can take the superpotential to have the form:
\beq
W = W_o + \alpha z + \beta z^2 + \gamma z^3 + \dots
\eeq
(we will see that terms higher than cubic are irrelevant not only
in the renormalization group sense, but in our analysis below), and a Kahler
potential:
\beq
K = a + bz + b^* z^* + c z^2 + c^* z^{*2} + d z^* z + \dots
\eeq
In perturbation theory, $a,...d \sim 1$, and we will assume that this
is general.

Now we want to impose the following conditions:
\begin{enumerate}
\item  $F$ is small at the minimum, i.e.
\beq
\alpha + b W_o =F~~~~\vert F \vert < F^*.
\eeq
\item  The potential is zero at the minimum.  Since $F$ is small, this means that $W_o \sim F$.
\item  The potential has its minimum at $z=0$.
Douglas and Denef provide a convenient formula for the derivatives of the potential:
\beq
\partial_z V = e^{K}(D_z D_z W \bar D^{\bar z} \bar W - 2 D_z W \bar W).
\eeq
\item  The potential is metastable.  Again, Douglas and Denef provide convenient formulas
for the second derivatives of the potential.  The relevant derivatives for us will be:
\beq
\partial_z^2 V = e^{K}(D_z D_z D_z W \bar D^{\bar z} \bar W - D_z D_z W \bar W)
\eeq
\end{enumerate}

If we assume that the Kahler potential is bounded, $b$ cannot become arbitrarily large.
So from the first two conditions we learn that $\alpha \sim F$.
Then the third condition gives that $\beta \sim F$.
Finally, if $\gamma$ is large, examining the second derivative terms, we see that the
masses cannot be positive; so once more, $\gamma \sim F$.

The distribution of $W_o$ is known in many examples to be uniform at small $W_o$\cite{Denef:2004ze}.
We will argue
in the next subsection that this is quite general, and also explain
when one expects exceptions.  So
assume that the distributions of $\alpha$, $\beta$ and $\gamma$ are uniform
at small $\alpha,\beta,\gamma$ as complex numbers.  The fraction of states with
suitable $\alpha$ is of order $\vert F \vert^2$; similarly for $\beta$ and $\gamma$.
The fraction with cosmological constant smaller than $\Lambda_o$ is of order $\Lambda_o$.
(Throughout this discussion we are assuming that apart from the separation of the
Planck scale from the supersymmetry breaking scale, there are no other large ratios
of scales, i.e. the string scale and four dimensional Planck scale are comparable.)
In other words, the number of states is of order
\beq
\int d^2 \alpha d^2 \beta d^2 \gamma d^2 W_o \theta(\Lambda_o -V) \theta(F^*-\vert
\alpha \vert) \theta(F^*-\vert
\beta \vert)\theta(F^*-\vert
\gamma \vert)
\sim \Lambda_o F^{*6}.
\eeq

These estimates appear to have a quite general character.  They rely only
on the assumption of roughly uniform densities in the various parameters of the low
energy superpotential.  In the next subsection,
we explain why this assumption is quite weak.
The general result accords quite closely with conventional ideas about
naturalness.  Phenomena on branches two and three, as we will see, are also
consistent with such expectations.

\subsection{Uniformity of the Measure on the Low Energy Parameters}

The assumption of uniform densities about a particular
point in the parameter space, $\alpha_o,\beta_o,\gamma_o,\dots$
is quite weak.  It follows if:
\begin{enumerate}
\item  The number of states near this point is sufficiently large that the distribution
is approximately continuous, i.e. there are many stationary points in this neighborhood.
\item  The distribution function has a Taylor series about the point.
\end{enumerate}

In our case, the assumption of uniformity is the assumption
that the point where supersymmetry breaking small is not distinguished.
In fact, what distinguishes branches two and three from
branch one is precisely the fact that various distributions
are not uniform\cite{Dine:2004is}, because assumption (2) does not hold.
If the origin of supersymmetry breaking is dynamical, then the distribution
of $F$'s behaves as
\beq
\int {d^2 F \over \vert F \vert^2}.
\eeq
In theories with unbroken supersymmetry at tree level, non-renormalization
theorems account for the special character of small $F$.  This sort of singular
behavior also occurs if the hierarchy arises through warping\cite{Kachru:2003aw}.
In the third branch of \cite{Dine:2004is}, the $W_o$ distribution also is singular
at the origin; this is connected with
the fact that the origin is a point of enhanced
symmetry ($R$ symmetry).  Coupled with the requirement
of small cosmological constant, it leads\cite{Dine:2004is}
to an even more singular distribution,
\beq
\Lambda_o\int {d^2 F \over \vert F \vert^4}.    
\eeq
It is these behaviors which distinguish the three branches.

\subsection{Stabilizing the Kahler Moduli}

The analysis of Douglas and Denef and the numerical experiments described in the
previous sections ignore the Kahler moduli.  Here we argue that, much as in the original
KKLT analysis, for small $F$ it is likely that the Kahler moduli are often stabilized
at large values of the radii.  Our effective lagrangian setup is particularly
convenient for this discussion.

Following $KKLT$, we simplify the analysis by considering only a single Kahler modulus,
$\rho = i \sigma + \alpha$.
We suppose that the superpotential has the form:
\beq
W = W(z) + e^{i c \rho}
\eeq
($W(z)$ is our earlier superpotential)
while we take for the Kahler potential,
\beq
K = - 3 \ln(i(\rho -\rho^*)).
\eeq
We look for a stationary point of the potential with
\beq
D_\rho W = {\partial W \over \partial \rho} + {\partial K \over \partial \rho} W
\eeq
This equation is particularly easy to analyze with the assumption that the solution
lies at large $\sigma$ and near $z=0$.  Then:
\beq
c e^{-c\sigma} \approx {3 \over 2 \sigma} W_o
\eeq
so
\beq
\sigma \approx - {1 \over c}\ln(W_o).
\eeq

This appears to be approximately self-consistent.
However, the mass
of the $\sigma$ field is not so small.
\beq
m_{\sigma}^2 \approx {1 \over \rho} \vert W_o \vert^2
 \approx \rho^2 m_{3/2}^2.
 \eeq
 The mass of the lightest $z$ field (associated with the small $F$) is of order
 $m_{3/2}^2$.  So one might
 worry that it is not entirely consistent to
first integrate out $z$ and then solve for $\rho$. 
The question is:  at our would-be solution, how large is the tadpole for $z$, i.e.
is the shift small.  It is, in fact, easy to see that the shift is of order
$1/\rho$.  Roughly, the potential has the form:
\beq
V(z,\rho) \propto {1 \over \sigma^3} \left [\left \vert -c e^{-c \sigma} + 3 {W_o+\alpha z + \beta z^2
+  \gamma z^3 \over \sigma}
\right \vert^2 \sigma^2 - 3 \left \vert W_o + \alpha z+ \beta z^2
+\gamma z^3 + e^{-c \sigma}\right \vert^2 \right ]
\eeq
As $D_\rho W$ vanishes to lowest order in $1/\sigma$, there is no contribution linear in
$z$ from the first term in the brackets, and the second term is suppressed by $1/\sigma$.

This argument suggests that plausibly all of the moduli can be fixed, with small
supersymmetry breaking, in a subset of vacua.

\subsection{Lifetime of Non-Supersymmetric States}

Perhaps the best reason to suspect the existence of a physical cutoff is the question
of stability.  In our simple model, the infinite set of states lies largely at very large
cosmological constant.  As we explain here (and as is not at all surprising), most of these
would-be states are extremely unstable.

To see this, recall again that the cosmological constant of these would-be
states is of order $N^2$, i.e. it is very large.  They can decay in various
ways.  Consider, first, decays of dS states to the region of large radius.
Simple scaling arguments indicate that the action for the bounce in a semi-classical
treatment scales as $1/N^2$, for large $N$.  In other words, the semiclassical
calculation breaks down already for $N=1$; one expects that the widths of the
states are order one or larger.  Alternatively, one can state this in
terms of supersymmetry breaking scales.  For a scale $F$, the
decay rate scales as $e^{-1/\vert F \vert^2}$.
As we noted in the previous expression,
once $N \gg 1$, all approximations are breaking down in the typical state;
there is no clear sense in which there are metastable states at all.

The actual scaling argument is quite simple.  For this we can follow Coleman
in the case without gravity.  Consider a potential of the form:
\beq
V = V_o f(\phi)
\eeq
for some scalar field $\phi$.  For an $O(4)$ symmetric bounce, the action
is
\beq
S = \int dr r^{d-1}((\partial_r \phi)^2 + V_o f(\phi))
\eeq
Now simply rescale $r= (V_o^{-1/2})u$. to see that the action scales
as
$V_o^{1-d/2}$.

For the subset of states with small cosmological constant, this particular decay
channel is not important, even for large supersymmetry breaking, and we have
seen that there are potentially many such states.

But
there are others which clearly must be considered.  Typically, if we consider
the space of possible fluxes, near the choice of fluxes which leads to a
state of small cosmological constant, there will be AdS states.  We have
not worked out the theory of such decays, but
it is presumably possible to decay with change of flux and
emission of topological defects.
As for ordinary vacuum decay, gravity surely effects these processes in important
ways.  Would-be decays to AdS states, for example, will sometimes not occur.
In cases where they can, the final configuration is likely to be a big crunch,
and the lifetimes, in general, short.  Understanding these sorts of considerations
is likely to bear on the question of the relative importance/likelihood of branch one.

\section{Statistics on Branch 2:  Are they computable?}

Our principle observation of the last section is that one cannot calculate
even the crudest statistics on branch one, the non-supersymmetric branch.
This raises the question:  can one actually do better on branch two (we will
comment on the case of branch three in the next section).

A major point of the KKLT paper was to exhibit supersymmetric AdS vacua with
all moduli fixed in regime in which the string coupling is small and the
compactification radii large.
In the KKLT analysis, large $\rho$ arises when one solves the equation
$
D_\rho W = 0$
with $W = W_o + e^{ic\rho}$.  But this form of $W$ is only valid for large $\rho$,
and there is no particular reason to think that there are not solutions for small $\rho$.
For such solutions, $<W>$ is presumably randomly distributed as a complex variable.
Other quantities, like the values of couplings, might also be randomly distributed.
A priori, it is not clear why these should not be at least as common as the solutions
with large $\rho$.

This suggests that if we ask about states at small $\langle W \rangle$ and
small $\Lambda$, the number with
small $\rho$ is likely comparable to and possibly significantly larger than
the number with large $\rho$ (we thank Shamit Kachru for stressing this
point to us).  Among these states could be states with spectra and
couplings similar to those we observe.

Reasonable assumptions can be used to argue that nature might be in the
large $\rho$ regime.  The observed gauge couplings are small.  In the Type IIB setup,
the (inverse) gauge couplings at large $\rho$ are proportional
to the value of $\rho$.  At small $\rho$ and strong string
coupling, it seems
plausible that they will be more or less uniformly distributed at small coupling.
But remember we want to account for small $\langle W \rangle$
and small couplings.  In the large $\rho$
limit, small couplings are a consequence of small $\langle W \rangle$.  In the small $\rho$ region,
we pay the same price for small $<W>$, and we pay an additional price ($10^{-6}$ perhaps?)
for small couplings.  Small couplings might well be selected by anthropic considerations,
or they may be a piece of data we wish to impose.

If we take coupling unification seriously, something like large $\rho$ looks even
more plausible.  We will not propose here a mechanism to understand even tree level
unification of couplings, but if there is some important class of states for which
unification is common, it is surely more likely to occur in the large $\rho$ region
(or its analog), then randomly in the strong coupling region.

\section{The $W=0$ Branch:  Discrete R symmetries}

In \cite{Dine:2004is}, it was argued that on the $W=0$
branch, small cosmological constant favors very low scale
breaking.  Roughly, the distribution behaves as:
\beq
{\cal N}(\vert F \vert < F^*,\Lambda < \Lambda_o) \propto \Lambda_o {1 \over F^*}.
\eeq
So, while on the one hand, one might expect the fraction of states with
unbroken $R$ symmetry might be small, there is a big gain when one selects
for the cosmological constant.  Further gains might arise from other
selection effects, such as proton decay.

Vanishing $W$ might arise by accident, or as a result of symmetries.  Ref. \cite{DeWolfe:2004ns}
performed some counting in simple models.  As one might naively guess, in most cases,
these authors found vanishing $W$ results from discrete R symmetries.  It is not hard to see
that such a symmetry is likely to yield both unbroken supersymmetry and vanishing $W$.  Suppose
that one has a discrete R symmetry under which
$W$ transforms by a phase $\alpha$.  Suppose
that there are some number of fields, $Z_i$, $i=1,\dots,n$, which also transform by $\alpha$,
and some number, $\phi_A$, $A=1,\dots ,m$, which are neutral.  Then the superpotential
has the form:
\beq
W= \sum_{i=1}^n Z_i f_i(\{\phi_A \}).
\eeq
Then
provided $m>n$, there will be supersymmetric solutions
with $W=0$.  There will not be supersymmetric solutions if
$m<n$.  One can readily modify this argument in the case that there
are various fields which transform with some other power of $\alpha$.

In the landscape, it is easy to see that
states with $m>n$ are statistically
significantly favored. In the Calabi-Yau case, there is a pairing
of complex structure moduli and fluxes.  Loosely, in order that
the low energy theory respect the symmetry, we expect that for any complex
structure modulus which transforms under the symmetry, we must
set the corresponding flux to zero.  So in order to have a large
number of states, and a small number of vanishing fluxes, we must
have a small number of fields which transform under the symmetry.
So we are in the limit $m>n$, above.  Of course, without a complete
survey, we cannot make a sweeping statement, but it seems likely
that $R$ symmetries in the landscape will lead typically to unbroken supersymmetry
and vanishing $W$.

It is helpful to illustrate these considerations with some examples.
One case where it is possible to be very explicit is that of
$IIB$ theory
on a $T_6/Z_2$ orientifold.  Focus on the point in the moduli space where the
$T_6$ is a product of three two-tori of equal size.  In this case, before turning
on fluxes, the theory has a variety of $R$-symmetries.  There are $Z_4$
symmetries which rotate each of the separate planes; there is also an $S_3$
symmetry which permutes the planes.  We can attempt to turn on only fluxes which
respect the $Z_4$ symmetry in the first plane.  Starting from the list in \cite{Kachru:2002he}
we can make a complete list:
From the subset:
\beq
\alpha_{ij}= {1 \over 2} \epsilon_{ilm}dx^\ell
\wedge dx^m \wedge dy^j~~~~~
\beta_{ij}= -{1 \over 2} \epsilon_{jlm}dy^\ell
\wedge dy^m \wedge dy^i~~~~~
\eeq
$\alpha_{12}$,$\alpha_{13}$, $\alpha_{21}$, $\alpha_{31}$
and similarly for the components of $\beta$.   For simplicity,
take for the RR and NS three-forms, $F$ and $H$:
\beq
{1 \over (2\pi)^2 \alpha^\prime} F_{(3)}
   = a^{12}\alpha_{12} + b_{12}\beta^{12} ~~~~~
   {1 \over (2\pi)^2 \alpha^\prime} H_{(3)}
   = c^{12}\alpha_{12} + d_{12}\beta^{12}
\eeq
Correspondingly, the superpotential is:
\beq
(a^{12}-\phi c^{12}) {\rm cof}\tau_{12}
   + (b_{12}-\phi d_{12})\tau^{12}.
   \eeq
We look for a solution of the form
\beq
\tau = \left ( \matrix{i & 0 & 0 \cr 0 & i & 0 \cr 0 & 0 & i} \right )
\eeq
Then $W=0$, and one finds for $\phi$:
\beq
\phi = {i a^{12} - b_{12} \over i c^{12} - d_{12}}
\eeq

Two remarks about this solution are important.  First, there is a multiparameter
set of solutions.  Many of the elements of $\tau$ are undetermined.
Second, we should also impose the tadpole cancellation condition.  In this
case, this reads:
\beq
-a^{12} d_{12} + b_{12}c^{12} = L
\eeq
with $L \le 16$ (and integer).
Restricting $\phi$ to the fundamental region leaves a limited set of fluxes.
Note, in particular, that if $L=1$, $\phi$ is an $SL(2,Z)$ transform of $i$.
There are hundreds of such solutions.  Additional solutions are obtained by
permuting the planes, and considering points with other symmetries, such as
$Z_6$.  This model is not realistic at many levels.  The vacua which we have
enumerated here, as pointed out in \cite{Kachru:2002he} have more than four
supersymmetries, for example.  Still, they exhibit many of the features
we discussed above.

In these cases, the superpotential transforms with a phase, $\alpha= e^{2\pi i \over 4}$;
the fields
$\tau^{12}$ and $\tau^{13}$ also transform with phase $\alpha$.  The only other
field which transforms non-trivially is $\tau^{11}$.  Because the superpotential
in this model is at most cubic, $\tau^{11}$ cannot appear and is thus not determined.
The rest of the superpotential has the structure:
\beq
W = \tau^{12} f(\tau^{22},\tau^{23},\tau^{33},\phi)
    + \tau^{13} g(\tau^{22},\tau^{23},\tau^{33},\phi).
\eeq

With sufficiently general choices of fluxes, one easily breaks all of the R symmetries.
Still, in this model, one often finds $W=0$\cite{DeWolfe:2004ns}.  This may be
a feature of this class of superpotentials, which are rather restricted polynomials,
but it might also be more general.  This is an important question to investigate.

Now consider the case of a more complicated Calabi-Yau space.  The quintic
in $CP^4$ is well studied, and its symmetries are explained in textbooks\cite{gsw}.
In particular, there is a highly symmetric point with symmetry $Z_5^4$ (times
a permutation symmetry, which we will ignore).  Calling the coordinates
of $CP^4$ $Z_i$, $i=1,\cdots,5$, the quintic polynomial equation which defines the
space is:
\beq
\sum Z_i^5 = 0.
\eeq
This is invariant under the separate symmetries $Z_i \rightarrow \alpha Z_i$,
where $\alpha$ is now a fifth root of unity.  The complex structure moduli are
in one to one correspondence with quintic polynomials.  Now we can require that
the fluxes preserve just one of the $Z_5$'s, say that acting on the
first coordinate.  Then the fluxes which are paired with polynomials (30 in all)
which transform under this $Z_5$ must vanish; examples include $Z_1^3 Z_2^2$,
$Z_1^3 Z_2 Z_3$, and so on.  So, in fact, there are 46 $Z$-type fields, in our
previous notation, and $55$ $\phi$ type fields.
One can try and preserve more $R$ symmetries.  Then many more fluxes must vanish -- and
there are many fewer states.  With a sufficiently large symmetry, there will
only be supersymmetry breaking solutions.

In this example, the dimensionality
of the flux space is reduced by more than 1/3.
While this is a dramatic reduction, there may be other selection effects which favor
such states.  The cosmological constant
itself is one; compared to the non-supersymmetric
case, this could well be a factor of $10^{50}$ or more; compared to branch two,
it could easily be $10^{25}$.  Proton decay is an additional
issue; without symmetries, on branch two, anthropic reasoning alone
cannot explain the value of the proton lifetime, and the anthropic constraint
itself requires a huge suppression ($10^{10}$ in several
couplings).   There may be other important selection effects as well.
How general this rather drastic reduction in the dimensionality of the space -- and, as
a result, in the number of states -- is a straightforward
problem worthy of further study.

Finally, we need to ask about those moduli which are unfixed at the level of the
superpotential equations.  These will be fixed by the dynamics which breaks
supersymmetry.  In general, there is no reason that these effects should
be calculable.  There are, potentially, severe cosmological problems associated
with these fields.  At the same time, this is a framework in which axions
might well arise, and interesting, testable long range forces (we thank
Savas Dimopoulos for stressing this possible positive aspect of these
fields to us).  We will comment further on these possibilities in the next section.

\section{Prospects for Predictions}

Of the three branches of the flux landscape which we have distinguished,
the first, non-supersymmetric branch is the most problematic.  We have seen
that, thanks to the work of Douglas and Denef\cite{Denef:2004ze}, the existence of this
branch is as well established as any branch of the landscape (explicit constructions
of states on branch one appear in \cite{Saltman:2004sn}).
We have given a simple argument based on the assumption
of uniform distribution of couplings in the low
energy effective action which reproduces their
results.  But the same arguments
which establish the statistics of these
states for small $F$ demonstrate that the vast majority of states
in this branch are inaccessible.  While within the set of states which can
be reliably studied, the number of supersymmetric and non-supersymmetric
states are comparable, we can only conjecture whether there are
or are not far more non-supersymmetric
states than supersymmetric ones, or, for that matter,
whether the
number of states on this branch is finite.  Even if we ignore cosmological
issues and simply assume that we should compare the relative numbers of states,
we cannot decide whether or not this branch dominates.  We have indicated
some simple cosmological questions which might affect this question.

If nature does lie on this branch, the situation is very disappointing.  The problem
is not that one does not predict low energy supersymmetry, but that one is
very unlikely to predict anything.  Because of the lack of control, there is
no underlying small parameter and approximation scheme which will allow one
to extract correlations of any kind.  The parameters of low energy physics
then presumably arise anthropically or randomly.

While one is largely in the dark about branch one, an optimist might proceed by
studying statistics at small $F$, hoping that some features will persist into the
inaccessible dominant region.  One question which one can address is how
hierarchies arise.  The work of \cite{Denef:2004cf,Giryavets:2004zr} indicates that warping is common
in the supersymmetric vacua.  This analysis remains valid in the almost supersymmetric
case, when there are
many moduli.  This feature might survive into the strong coupling regime.  An alternative
origin for hierarchies could be conventional dynamical supersymmetry
breaking, as in technicolor theories.  Just as in the case of branch
two, one expects that chiral gauge theories are common, and associated
with these dynamical symmetry breaking and
hierarchical scales.  Of course, problems
of precision electroweak physics, flavor and the like will all require some
resolution.  And the strong coupling problem means that it is probably impossible to
say more than what effective field theorists have argued for many years.

There are some phenomenological reasons to think we might
not be on branch one.  As we have just said, it would be necessary
to understand why solutions of the usual problems of
technicolor and/or warped spaces are solved in some
generic fashion.  More generally, in this strongly coupled regime,
the parameters of low energy physics would seem to be
either random or anthropic.  While it is plausible that several of the parameters
of the Standard Model are anthropic, it is implausible that most of them are.
There are clear patterns and regularities.  The small value of $\theta_{qcd}$
is one example\cite{Banks:2003es,Donoghue:2003vs}.  The heavy quark and lepton masses and mixings do not
appear random, yet they are probably not anthropic; the neutrino masses are much
larger than one might expect to arise
randomly in such a picture, and again there is no
obvious anthropic explanation.  (Of
course, these may be arguments that not only are we not on
branch one but that the landscape itself is not correct.)

One example of a prediction which is {\it not} likely to arise on
branch one is split supersymmetry
with very large scale of supersymmetry breaking\cite{Arkani-Hamed:2004fb}.  Small masses
for all the gauginos are probably not required anthropically.  One needs
broken supersymmetry at the high scale, but an unbroken $R$ symmetry.
But we have given simple arguments that in the landscape, unbroken $R$
symmetry is necessarily associated with unbroken supersymmetry -- in fact,
that this puts one on branch three.  A form
of split supersymmetry with a scale of order $10$'s of TeV's might arise for
other reasons\cite{Dine:2004is}.  But while the phenomenology
of much larger breakings is interesting to
to explore, it
it is not a generic feature of states of the landscape.

On branches two and three, it is more likely that one can make real predictions,
and resolve some of the puzzles of the Standard Model.  The number of states
appears to be finite, and the statistics seem compatible with conventional
notions of naturalness.  In \cite{Dine:2004is}, some features of branch two
and three were noted.  Here we looked mainly
at the relative populations
of branches two and three.  We adopted the point of view that vanishing $W$ arises,
in general, as a result of discrete $R$ symmetries.  This viewpoint finds some
support in the work of \cite{DeWolfe:2004ns}, who, while finding some cases
where vanishing $W$ arises by accident, largely found that the explanation
lies in such symmetries.  We explained why discrete $R$ symmetries are likely
to lead to both unbroken supersymmetry and unbroken $R$ invariance.   We also
explained why some moduli are likely to be unfixed in these models, before
non-perturbative effects, and particularly supersymmetry breaking, are taken
into account.  This is potentially problematic for cosmology, but also could be
a striking prediction\footnote{We thank Savas Dimopoulos for stressing this aspect
of light moduli to us.}

We asked the fraction of states which have vanishing $W$ at tree level in
a particular example with a large number of moduli.  In one example,
we found that requiring an
R symmetry reduced the dimension of the moduli space by about 1/3.  This is
presumably an enormous price in landscape terms, quite possibly larger than the
gain (about $10^{40}$) which one might imagine comes from the ease of obtaining
a small cosmological constant.  On the other hand, discrete symmetries might
be required by other considerations as well:  proton stability and dark matter.
These could easily introduce additional compensating factors of $10^{40}$ or so.

Making further progress on the phenomenology of branches two and three requires
more knowledge of the distribution of gauge groups and chiral matter content;
some preliminary
studies of these questions have been reported\cite{Kumar:2004pv,Conlon:2004ds,Blumenhagen:2004xx}.
More information
about statistics of discrete symmetries is also essential.
It would seem straightforward to
explore these questions.  With this sort of data, we could, for example,
assess the likelihood of low energy dynamical supersymmetry breaking
in branches two and three.  As explained in \cite{Dine:2004is}, if gauge
singlets are not common, on branch two one is lead to a prediction which in fact coincides with
one version of split supersymmetry.  On branch three, there are a rich set of
questions one might hope to address.  We have begun some preliminary studies
of these statistics.

Finally, the observations of this paper
suggest a more refined (perhaps we should
say mature) view of naturalness.  In the landscape,
conventional ideas of naturalness often have
a sharp realization.  Our success in reproducing the
formulas of Douglas and Denef is an example.  Enumerating the
terms in a low energy effective action, and assuming
uniform distributions for them, yields
the results of DD's more sophisticated microscopic
calculation.  Our arguments about
physics in the different branches, and about the relevant populations of
branch two and three are similar.  But we are presently stymied in our
efforts to decide about the population of, and importance of,
branch one relative to branches two and three.  Taking the cutoff in the DD calculation
to be order one makes it plausible that branch one and two
have comparable populations; the work of Silverstein
and collaborators\cite{Saltman:2004jh} and of Bobkov\cite{Bobkov:2004cy},
suggesting that there are numerous
other constructions of non-susy vacua, might suggest
that the population of branch one is
huge compared to that of branch two.
It is also possible that cosmological
considerations will ultimately force us to phrase the question in some very different
way.

From all of this, it appears that it is difficult, in principle, to decide
whether or not the landscape predicts supersymmetry.  If one hypothesizes
that it does, than one may be able to make predictions about the
scale and nature of supersymmetry breaking.  If it does not, it will be
difficult to extract any real predictions; at best, one might give a rationale
for the appearance of technicolor or warping in low energy physics -- or their
absence.

\noindent
{\bf Acknowledgements:}
We thank B. Acharya, T. Banks, F. Denef, M. Douglas, S.
Dimopoulos S. Kachru, G. Moore, S. Shenker, E. Silverstein, L.Susskind and S. Thomas
for sharing their insights and expertise on these questions.  We are particularly
grateful to Michael Douglas for critical reading of an earlier version of the manuscript.
This work supported in part by the U.S.
Department of Energy.

\end{document}